\documentclass[a4paper]{article}

\usepackage{INTERSPEECH2022}
\usepackage[skip=0pt]{caption}
\usepackage{hyperref}
\usepackage{graphicx}
\usepackage{subfigure}

\title{RefineGAN: Universally Generating Waveform Better than Ground Truth with Highly Accurate Pitch and Intensity Responses}
\name{Shengyuan Xu$^1$,Wenxiao Zhao$^2$,Jing Guo$^3$}
\address{
	timedomain.Inc}
\email{
	$^1$xushengyuan@timedomain.ai,
	$^2$sean.zhao@timedomain.ai,
	$^3$joe.g@timedomain.ai
}

\begin{document}
\setlength{\textfloatsep}{0pt}
\setlength{\floatsep}{0pt}
	\maketitle
	\footnotetext[1]{Submitted to INTERSPEECH2022}
	\begin{abstract}
		Most GAN(Generative Adversarial Network)-based approaches towards high-fidelity waveform generation heavily rely on discriminators to improve their performance. However, GAN methods introduce much uncertainty into the generation process and often result in mismatches of pitch and intensity, which is fatal when it comes to sensitive use cases such as singing voice synthesis(SVS). 
		
		To address this problem, we propose RefineGAN, a high-fidelity neural vocoder focused on the robustness, pitch and intensity accuracy, and high-speed full-band audio generation. We applyed a pitch-guided refine architecture with a multi-scale spectrogram-based loss function to help stabilize the training process and maintain the robustness of the neural vocoder while using the GAN-based training method. 
		
		Audio generated using this method shows a better performance in subjective tests when compared with the ground-truth audio. This result shows that the fidelity is even improved during the waveform reconstruction by eliminating defects produced by recording procedures. Moreover, it shows that models trained on a specified type of data can perform on totally unseen language and unseen speaker identically well. 
		Generated sample pairs are provided on \href{https://timedomain-tech.github.io/refinegan/}{https://timedomain-tech.github.io/refinegan/} .
		
	\end{abstract}
	
	\section{Introduction}
	
	\subsection{Conventional neural vocoders}
	Vocoders have been employed in various fields in counter with human voice recreation, such as text-to-speech(TTS), singing voice synthesis(SVS), and voice conversion. Moreover, vocoders are often the bottleneck when it comes to the voice quality. Recently a series of GAN-based methods\cite{kumar2019melgan,kong2020hifi} have been employed in vocoders to synthesize more realistic waveforms while also reducing synthetic artifacts in the voices. And these non-autoregressive CNN-based designs have increased the generation speed to an acceptable level.
	
	While most neural vocoders are focused on generating band-limited waveforms, a majority of replay devices natively support full-band audio. The quality-reduced audio delivered by those conventional vocoders are becoming increasingly unsuitable to be consumed. And audio quality is crucial when deep learning-based technology is applied to the workflow of content creating. Moreover, in neural network singing voice synthesis\cite{wu2020adversarially}, audio quality is considered as an obstacle constantly delays the technology from entering the existing business.
		
	\subsection{Towards finest waveform generation}
	We designed a way to use pitch guided methods to diminish the difficulty of generating voice waveform with higher sample rates. It makes stable, high-resolution voice synthesis affordable when using full-band Mel-spectrogram and the target pitch as network inputs. This approach is designed to make the task of conversion from the frequency domain to the time domain easier for the neural networks by using some simple external calculation.
	
	Conventional neural or non-neural\cite{morise2016world} vocoders center on eliminating most of the noticeable flaws and feature losses during the restoration. However, in our work, by using several methods to level down the learning difficulties, the amount of synthetic artifacts are highly reduced. Those defects are not audible to a large percentage of the content consumers or even listeners with professional experience in the audio industries. 

%
		
	\subsection{Related Works}
		
	Hifi-GAN\cite{kong2020hifi} introduced a unique multi-scale and multi-period discriminator design to produce high-fidelity voices. Evaluations on it had shown better audio quality and improved voice naturalness. Also, UnivNet\cite{jang2021univnet} presented a new discriminator design based on the idea of deciding on a linear spectrogram rather than on the waveform. This dramatically enhance the discriminator's performance and improve the audio quality in higher frequencies.
	\begin{figure}[t]
		\centering
		\includegraphics[width=180px]{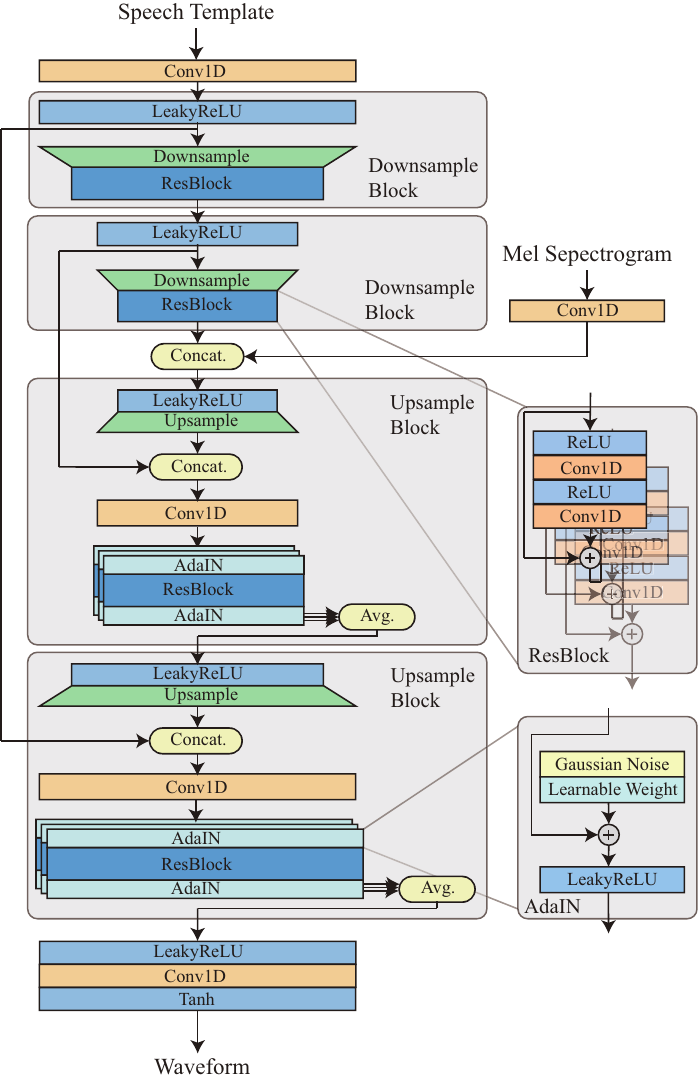}
		\caption{A figure showing the whole generator architecture}
		\label{fig:generator}
	\end{figure}
	
	\section{RefineGAN}
	\subsection{Generator}

	As we call it RefineGAN, we use an encoder-decoder-like architecture to refine a signal(as we call it, speech template) constructed from pitch information into the final speech waveform. 
	
	The refinement process includes an encoder module to encode the speech template into a intermediate form to apply Mel-spectrogram information into it, and a decoder module to generate the waveform from this hidden form. 
	The encoding and decoding processes are done by first downsampling the waveform into a shape the same as the Mel-spectrogram and then upsampling it back into the final waveform. The downsampling process is done by using convolution layers and corresponding transposed convolution layers for the upsampling process.
	
	To make pitch information fully accessible, we also add a cross-connection mechanism between each block in the encoder and decoder parts, using an UNet\cite{ronneberger2015u}-like strategy. We employ the design of using parallel ResBlocks\cite{he2016deep} with differed kernel sizes and dilation sizes, which is similar to the architecture of Hifi-GAN. Each of these ResBlocks contains three sub-blocks connected in series. Furthermore, in each sub-block, inputs are passed through 2 weight normalized convolution layers with LeakyReLU\cite{xu2015empirical} pre-applied.

	\subsection{Speech template}
	
	We generate the speech template from the target pitch information using methods shown below. The speech template signal is the same length as the target waveform and uses the same sample rate. First, we obtain the pitch information of the signal by using Harvest\cite{morise2017harvest} algorithm. For unvoiced parts in the target signal, we generate uniform noise as the noise source of the GAN network. 
	\begin{figure}[htb]
		\centering
		\includegraphics[width=140px]{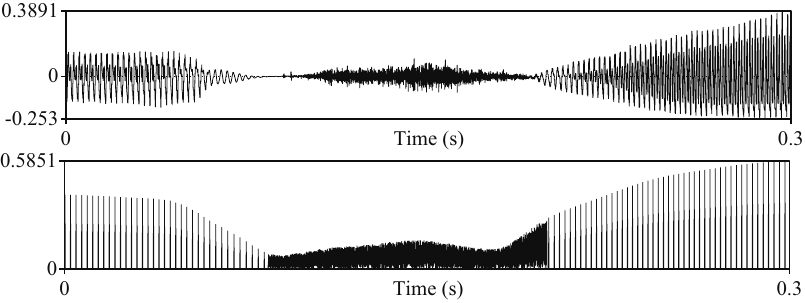}
		\caption{A comparison between speech template and waveform generated from it in a large scale.}
		\label{fig:pulse_and_audio_large}
	\end{figure}
	
	As showed in Figure \ref{fig:pulse_and_audio_large}, for voiced parts, we generate one-sample-long pulses, and the time between each pulse is calculated from the target pitch as the reciprocal of the frequency at the corresponding time. We define the values of the pulses as intensity-like values calculated from the Mel-spectrogram.
	\begin{figure}[htb]
		\centering
		\includegraphics[width=140px]{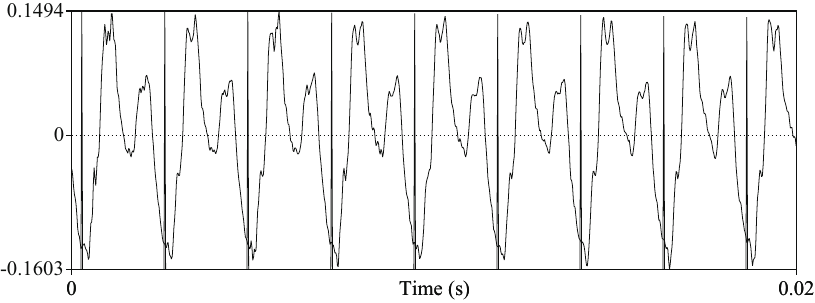}
		\caption{The expecting behavior for the model to fill the signal between pulses under the guide of the speech template.}
		\label{fig:pulse_and_audio_small}
	\end{figure}
	A speech template like this acts as a vital element to significantly lower the training difficulties. Because it directly gives the model the information of the exact position and the precise length of each pulse in the target signal. Then as showed in Figure \ref{fig:pulse_and_audio_small} the model merely needs to learn to appropriately fill the pulse signal into each blank between pulses instead of directly generating all the features from the ground.

	\subsection{Pitch-based data augmentation}
	For a broader coverage when using a relatively small dataset, we aim to randomly generate training data with different voices. So we perform a random pitch shift on the utterances randomly sliced from the source recordings.
	We uniformly choose the amount of pitch shifting $\zeta$, which is measured in semitones, from lower limit $\zeta_\text{min}$ to upper limit $\zeta_\text{max}$.
	
\begin{figure}[ht]
		\centering 
		\includegraphics[width=140px]{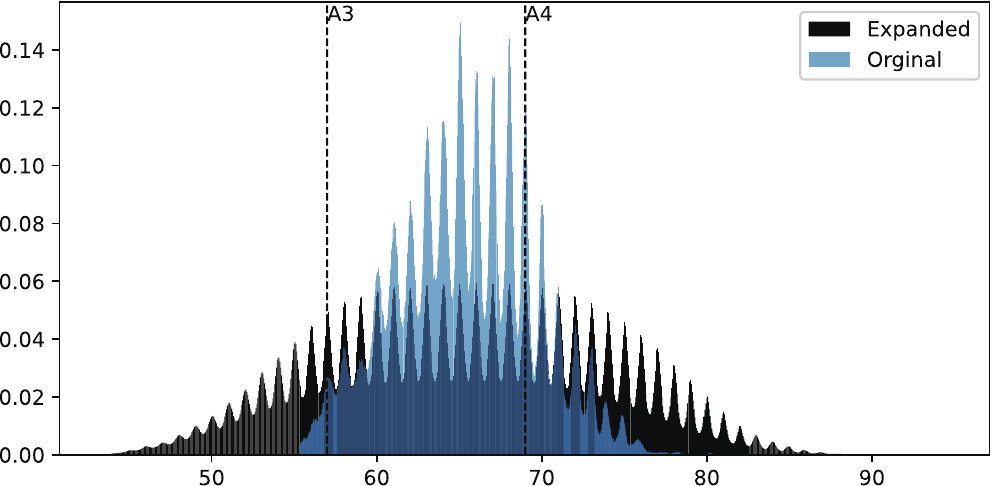}
		\caption{A histogram showing the pitch distribution of the original recordings from one of the female singer and the extended audio when we set $\zeta_\text{min}=-12 , \zeta_\text{max}=12$. (Pitch is measured in semitones. A4=69=440Hz)}
		\label{fig:pitch_hist}
	\end{figure}

	To avoid any quality losses and introduced artifacts while shifting the pitch, we modify the length of the audio signal simultaneously, so all of this can just be efficiently done by performing a resampling algorithm. A simple and computation-friendly approach to this target enables us to generate training data dynamically when training. As showed in Figure \ref{fig:pitch_hist} this method provides data with an extremely high or low pitch to train the model on these rare circumstances.
			
	\subsection{Multi-param Mel-spectrogram loss function}
	When extracting Mel-spectrograms from predicted signal and the ground-truth audio, we cannot achieve the maximum accuracy representing features both in the time domain and the frequency domain at the same time with the same FFT size. To comprehensively evaluate the difference between two signals, we chose distinctive sets of parameters on individual levels to compute the Mel-spectrograms. Then we use the average value of the MSE losses between the predicted signal and the ground truth as the final Mel-spectrogram loss.
	\begin{equation}
		\mathcal{L}_{\text {mel}}(\mathbf{y},\mathbf{\hat{y}}) = \frac{1}{n} \sum_{i=1}^{n} \lVert\log \it{M}_i(\mathbf{y})-\log \it{M}_i(\hat{\mathbf{y}})\rVert
	\end{equation}
	where $n$ is the numbers of parameter sets and function $\text{M}_i$ represents calculating Mel-spectrogram from waveform using the set of parameters $i$.
	
	\subsection{Loudness focused enhancements}
	\subsubsection{Envelope loss function} 
	To further enhance the intensity control of the vocoder, we extract the envelope feature from the signal by using a 1D max-pooling layer and add a envelope-based loss function. We apply max-pooling layers both to the original audio signal and to the audio with reversed polarity to get the envelope curves. Then we calculate the sum of the MAE losses of the envelopes :
	\begin{equation}
		\mathcal{L}_{\text {envelope}}(\mathbf{y},\mathbf{\hat{y}}) =  \lvert\text{pool}(\mathbf{y})-\text{pool}(\mathbf{\hat{y}})\rvert
		 +\lvert\text{pool}(-\mathbf{y})-\text{pool}(-\mathbf{\hat{y}})\rvert
	\end{equation}

	\subsubsection{Loudness-based data augmentation}
	To make it possible for the model to generate a high-quality waveform no matter the loudness level of the utterance, we also applied a loudness-based data augmentation technique. This approach allows randomly adjusting the loudness level of each utterance within an acceptable range. We control the range by defining four parameters: $p_\text{min}$, $p_\text{max}$, which indicates the audio signal's minimum and maximum peak value; $r_\text{min}$, $r_\text{max}$, which indicates the minimum and maximum rate of the loudness adjustments. We can randomly sample the target peak value :
	\begin{equation}
		p' \sim \text{U}[\max(p_\text{min}, r_\text{min}p), \min(p_\text{max}, r_\text{max}p)]
	\end{equation}
	where $p$ means the original peak value $\max(|y|)$.
	
	And we can apply the gain to the audio signal by simply calculating $y' = \frac{y p'}{p}$.
	where $y$ and $y'$ represents the audio signal before and after the augmentation process.

	In this case, we did a uniform random sampling on a linear scale. And we can train the model with a much greater loudness diversity and achieve a state where the model's performance will not be affected by the loudness level.
	
	\subsection{Discriminator}
	We use two set of discriminators to bring a broader coverage of various features. One set of them is the same Multi-Period Discriminator we adopt from Hifi-GAN\cite{kong2020hifi}. We use period parameters of 2,3,5,7,11 in the hope of being better suitable for 44100hz full-band generation. The other is the Multi-Resolution Discriminator we adopt from the UnviNet\cite{jang2021univnet} and use the same three different sets of parameters as the author mentioned in the paper.
	
	\subsection{Training losses}
	\subsubsection{Generator loss}
	For the generator, we add up 3 different loss elements we mentioned above to construct the final generator loss. And also add a weight $\lambda$ to the $\mathcal{L}_{\text {mel}}$ to help stable the training process:
	\begin{equation}
		\begin{aligned}
			\mathcal{L}_{\text {G}} = & \lambda \mathcal{L}_{\text {mel}}(\mathbf{y},\it{G}(\mathbf{z},\mathbf{c}))) + \mathcal{L}_{\text {envelope}}(\mathbf{y},\it{G}(\mathbf{z},\mathbf{c})) \\
			& +\frac{1}{n_{\text{MPD}}} \sum_{i=1}^{n_{\text{MPD}}}  \mathbb{E}_{\mathbf{z},\mathbf{c}} [ \log ( 1+ \exp(-\it{D}_{\text{MPD},i}(\it{G}(\mathbf{z},\mathbf{c})))) ] \\
			& +\frac{1}{n_{\text{MRD}}} \sum_{i=1}^{n_{\text{MRD}}}  \mathbb{E}_{\mathbf{z},\mathbf{c}} [  \log ( 1+ \exp(-\it{D}_{\text{MRD},i}(\it{M}_i(\it{G}(\mathbf{z},\mathbf{c})))))]
		\end{aligned}
	\end{equation}
	
	\subsubsection{Discriminator loss}
	Within both the Multi-Period Discriminator and the Multi-Resolution Discriminator, we use an average value of the output from modules using different parameters and a single Adam optimizer for both the discriminators. And as described below, the sum of the MPD's and MRD's output acts as the loss for backpropagation:
	\begin{equation}
		\begin{aligned}
			\mathcal{L}_{\text {MPD}} =& \frac{1}{n_{\text{MPD}}} \sum_{i=1}^{n_{\text{MPD}}} ( \mathbb{E}_\mathbf{y} [\log ( 1+ \exp(-\it{D}_{\text{MPD},i}(\mathbf{y}) )) ]\\
			 &+\mathbb{E}_{\mathbf{z},\mathbf{c}} [ \log ( 1+ \exp(\it{D}_{\text{MPD},i}(\it{G}(\mathbf{z},\mathbf{c})))) ])
		\end{aligned}
	\end{equation}
	\begin{equation}
		\begin{aligned}
			\mathcal{L}_{\text {MRD}} =& \frac{1}{n_{\text{MRD}}} \sum_{i=1}^{n_{\text{MRD}}}  ( \mathbb{E}_\mathbf{y} [\log ( 1+ \exp(-\it{D}_{\text{MRD},i}(\it{M}_i(\mathbf{y})))) ]\\
			&+\mathbb{E}_{\mathbf{z},\mathbf{c}} [  \log ( 1+ \exp(\it{D}_{\text{MRD},i}(\it{M}_i(\it{G}(\mathbf{z},\mathbf{c})))))] )
		\end{aligned}
	\end{equation}
	\begin{equation}
		\mathcal{L}_{\text {D}} = \mathcal{L}_{\text {MPD}} + \mathcal{L}_{\text {MRD}}
	\end{equation}
	\section{Experiments}
	
	\subsection{Hyperparameters}
	We select 44100Hz as the sample rate to achieve the goal of full-band voice generation. Furthermore, to extract the Mel-spectrogram as the condition, we choose 128 as the number of the Mel channels. Then chose 2048 as both the FFT size and the windows size while 256 as the hop size. 
		
	In our model, we use each four layers in the encoder part and the decoder part. Between layers, we use downsampling rates of 2,2,8,8 on the encoding side and correspondingly 8,8,2,2 on the decoding side, and the kernel sizes of the these layers are all set to 2 times the rates. All the ResBlocks use the exact dilation sizes of 1,3 and 5, but parallel ResBlocks with different kernel sizes are only applyed in the decoding side with kernel sizes of 3,7 and 11. In the encoding side, only one ResBlock with a kernel size of 7 are employed in each layer to reduce the model size.

	\subsection{Training}
	As we focused on Singing Voice Synthesis(SVS), we trained the model entirely using studio-grade quality singing recordings with a length of 59 hours from 12 different Chinese singers, the amounts of the data are shown in Table \ref{tab:singing data}. In training, we randomly select slices from all the audio and calculate the Mel-spectrogram and the speech template after data augmentations. 
	
	Calculations of the conditions are all done simultaneously with the training process. With our data augmentation methods mentioned above, this approach formulates a great diversity of the training data and significantly strengthens the robustness of the model.
	\begin{table}[th]
		\caption{Lengths and numbers of recordings of the singing voice dataset}
		\label{tab:singing data}
		\centering
		\begin{tabular}{cccc}
			\toprule
			&\textbf{Singers} & \textbf{Length(Hours)}  &\textbf{Recordings}  \\
			\midrule
			Female Singers &8	&41.68	&2325\\
			Male Singers &4  &18.08	&697\\
			All	&12 &59.77	&3022\\
			\bottomrule
		\end{tabular}
	\end{table}
	
\begin{table*}[htb]
	\caption{MOS test result on speech and singing voice in three languages}
	\label{tab:multi_lang_mos}
	\centering
	\begin{tabular}{cccccc}
		\toprule
		& \textbf{Griffin-Lim}  &\textbf{Hifi-GAN} &\textbf{Univnet}  &\textbf{RefineGAN (ours)} &\textbf{Ground Truth} \\
		\midrule
		Chinese speech	&1.45 $\pm$ 0.05  &3.19 $\pm$ 0.03 &2.79 $\pm$ 0.04 &\textbf{4.33 $\pm$ 0.02} &4.42 $\pm$ 0.02\\
		Chinese singing	&1.40 $\pm$ 0.05  &2.71 $\pm$ 0.04 &2.53 $\pm$ 0.04 &\textbf{4.34 $\pm$ 0.02} &4.33 $\pm$ 0.03\\
		Japanese speech	&1.45 $\pm$ 0.05  &3.14 $\pm$ 0.04 &2.95 $\pm$ 0.04 &\textbf{4.29 $\pm$ 0.02} &4.19 $\pm$ 0.03\\
		Japanese singing &1.42 $\pm$ 0.05  &2.76 $\pm$ 0.04 &2.51 $\pm$ 0.04 &\textbf{4.25 $\pm$ 0.02} &4.29 $\pm$ 0.03\\
		English speech	&1.61 $\pm$ 0.04  &2.72 $\pm$ 0.04 &2.46 $\pm$ 0.05 &\textbf{3.42 $\pm$ 0.04} &4.06 $\pm$ 0.03\\
		
		\bottomrule
	\end{tabular}
\end{table*}
	\section{Results}
	\subsection{General Performance}
		
	We trained the model on 8 RTX3090 GPUs for 450k steps with a training time of approximately 130 hours using a batch size of 16. An on a single RTX3090 we can achieve a inference speed of approximately 110X real-time. And further experiments show that we can easily convert the whole model into FP16 (without using automatic mixed precision) in order to get benefits from optimized hardware with nearly no audible quality loss. Generated samples are available on our web page.

	We also conducted a MOS test on 73 listeners. This test compares the subjective scores of ground-truth recordings, audio reconstructed by the model, and audio reconstructed by the Griffin-Lim\cite{griffin1984signal} algorithm. Each of the listeners scored 30 different 23 second-long utterances, and the result is shown in the Table \ref{tab:seen_speaker_mos}. 
	
		\begin{table}[th]
		\caption{MOS test result on seen speakers}
		\label{tab:seen_speaker_mos}
		\centering
		\begin{tabular}{cccc}
			\toprule 
			& \textbf{Griffin-Lim}  &\textbf{Predicted} &\textbf{Ground Truth} \\
			\midrule
			Male	&1.93 $\pm$ 0.09&\textbf{3.68 $\pm$ 0.09}&3.67 $\pm$ 0.09\\
			Female	&2.15 $\pm$ 0.10	&\textbf{4.02 $\pm$ 0.09}&3.97 $\pm$ 0.09\\
			All	&2.04 $\pm$ 0.07	&\textbf{3.85 $\pm$ 0.07}&3.82 $\pm$ 0.07\\
			\bottomrule
		\end{tabular}
	\end{table}
	
	\subsection{Performance on unseen speakers, unseen languages, and speech signals}
	
	One of our goals is to achieve universal waveform regeneration, so we studied using this trained model to predict singing voices in unseen languages and with unseen speakers. Supplemental samples are also provided on the web page. After analysis on more samples, we can conclude that our model can perform on unseen data nearly identically well as on the seen data.
	
	We also studied using the model trained with only singing voice recordings to reconstruct speech signals. We experimented on reconstructing speech in seen and unseen languages, and apart from Chinese\cite{AISHELL-3_2020}, English\cite{bakhturina2021hi}, French\cite{lemoine:hal-02508362} and Japanese\cite{sonobe2017jsut}, we also evaluated the performance on some under-resourced languages mentioned in \cite{kjartansson-etal-tts-sltu2018}. In these provided pairs we can see that even when merely trained with Chinese singing data, the model can still perform well on speech signals in even unseen language.
	\begin{table}[t]
		\caption{Recording lengths of the extended dataset. 
			*Including under-resourced languages and accents provided in \cite{kjartansson-etal-tts-sltu2018}}
		\label{tab:extended_data}
		\centering
		\begin{tabular}{cccc}
			\toprule
			\textbf{Type} & \textbf{Length} &\textbf{Type} & \textbf{Length}  \\
			\midrule
			Chinese singing	&59.77	&Chinese speech &47.38\\
			Japanese singing &2.66	&Japanese speech &9.06\\
			English speech	&5.83 	&French speech & 10.06\\
			Mixed speech* &27.69&&\\
			\textbf{All}	&\textbf{162.45}	&&\\
			\bottomrule
		\end{tabular}
		
	\end{table}

	\section{Further Study}
	\subsection{Traning with a comprehensive dataset}
		
	Interested in the outperforming generalization capabilities shown in the experiments above, we further studied the model on a highly diverse dataset. Table \ref{tab:extended_data} shows the composition of the dataset which includes studio-grade quality voices of different types(singing and speech) in servals languages from over 150 speakers. And a mixed dataset like this usually presents enormous difficulties for conventional neural vocoders' convergence. We use the same hyperparameters as above and trained the model on the same setup for 450k steps with a batch size of 16.
	
	Also we selected Hifi-GAN\cite{kong2020hifi} and UnivNet\cite{jang2021univnet} for comparison and use the official implements to train the models on the same dataset. We use the same hyperparameters provided in the implements and trained to convergence with the same 8-card setup , which takes nearly 150 hours each.

	\subsection{Results}
	\begin{figure}
		\centering
		\subfigure[ground truth]{
			\begin{minipage}[b]{0.20\textwidth}
				\includegraphics[width=1\textwidth,height=40px]{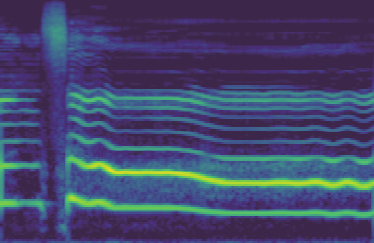} 
			\end{minipage}
			\label{fig:gt}
		}
		\subfigure[refineGAN]{
			\begin{minipage}[b]{0.20\textwidth}
				\includegraphics[width=1\textwidth,height=40px]{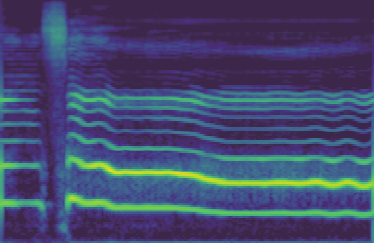}
			\end{minipage}
			\label{fig:refine}
		}
		\\ 
		\subfigure[Hifi-GAN]{
			\begin{minipage}[b]{0.20\textwidth}
				\includegraphics[width=1\textwidth,height=40px]{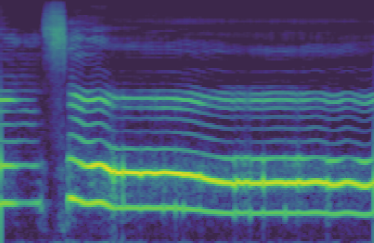} 
			\end{minipage}
			\label{fig:hifi}
		}
		\subfigure[UnivNet]{
			\begin{minipage}[b]{0.20\textwidth}
				\includegraphics[width=1\textwidth,height=40px]{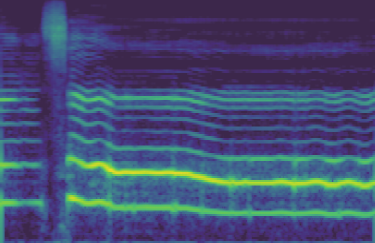}
			\end{minipage}
			\label{fig:univ}
		}
		\caption{Comparison of the waveform reconstructed by three methods. (Using a slice from a Chinese singing recording for demonstration)}
		\label{fig:mel_compare}
	\end{figure}
	We conducted a listening test on 67 listeners with the samples produced by these methods. and MOS results is showed below in Table \ref{tab:multi_lang_mos}. The result shows our approach has a much advanced ability of universially modeling voice signals regardless of the language, the speaker or even whether it is singing voice or speech voice. And as showed in Figure \ref{fig:mel_compare}, our method dramatically improved the pitch stability and solved the problem of generating pitch glitches when using conventional neural vocoders.

	\section{Conclusions}
	In this paper, we proposed RefineGAN, which consists of a series of methods to boost the performance of GAN-based neural vocoders. Parts of the methods mentioned above focused on improvements to the generator's architecture, while parts focused on improving training data's diversity. 
		
	The employment of these methods leads to model's outstanding robustness and stability, which enables the model trained on one particular type of data to be used to generate voices universally. Because the speaker or the language does not influence the voice quality much, this approach enables high-quality waveform synthesis even when the specific type of voice does not have sufficient high-fidelity recordings available for research.
		
	\newpage
	\bibliographystyle{IEEEtran}
	\bibliography{mybib}
		
\end{document}